\documentclass[aps,pra,twocolumn,superscriptaddress,longbibliography]{revtex4-2}
\usepackage{amsmath}
\usepackage{braket} 
\usepackage{url}
\usepackage{graphicx}
\usepackage{dcolumn}
\usepackage{bm}
\usepackage{amssymb}
\usepackage{rotating}
\usepackage[abs]{overpic}
\usepackage{xcolor}
\usepackage{tabularx}
\usepackage{hyperref}
\hypersetup{colorlinks = true, citebordercolor={blue}, linkcolor={blue}, citecolor={blue}, urlcolor={blue}}

\DeclareUnicodeCharacter{2212}{-}
\DeclareUnicodeCharacter{2009}{-}

\begin{document}



\title{Correlated Purification for Restoring $N$-Representability in Quantum Simulation}

\author{Yuchen Wang}
\affiliation{Department of Chemistry and The James Franck Institute, The University of Chicago, Chicago, IL 60637 USA}
\author{Irma Avdic}\thanks{These authors contributed equally to this work.}
\affiliation{Department of Chemistry and The James Franck Institute, The University of Chicago, Chicago, IL 60637 USA}
\author{Michael Rose}\thanks{These authors contributed equally to this work.}
\affiliation{Department of Chemistry and The James Franck Institute, The University of Chicago, Chicago, IL 60637 USA}
\author{Lillian I. Payne Torres}
\affiliation{Department of Chemistry and The James Franck Institute, The University of Chicago, Chicago, IL 60637 USA}
\author{Anna O. Schouten}
\affiliation{Department of Chemistry and The James Franck Institute, The University of Chicago, Chicago, IL 60637 USA}
\author{Kevin J. Sung}
\email{kevinsung@ibm.com}
\affiliation{IBM Quantum, IBM T.J. Watson Research Center, Yorktown Heights, NY 10598 USA}
\author{David A. Mazziotti}
\email{damazz@uchicago.edu}
\affiliation{Department of Chemistry and The James Franck Institute, The University of Chicago, Chicago, IL 60637 USA}

\date{Submitted November 13, 2025}

\begin{abstract}
Classical shadow tomography offers a scalable route to estimating properties of quantum states, but the resulting reduced density matrices (RDMs) often violate constraints that ensure they represent $N$-electron states---known as $N$-representability conditions---because of statistical and hardware noise. We present a correlated purification framework based on semidefinite programming to restore accuracy to these noisy, unphysical two-electron RDMs. The method performs a bi-objective optimization that minimizes both the many-electron energy and the nuclear norm of the change in the measured 2-RDM. The nuclear norm, often employed in matrix completion, promotes low-rank, physically meaningful corrections to the 2-RDM, while the energy term acts as a regularization term that can improve the purity of the ground state. While the method is particularly effective for the ground state, it can also be applied to excited and non-stationary states by decreasing the weight of the energy relative to the error norm. In an application to fermionic shadow tomography of large hydrogen chains, correlated purification yields substantial reductions in both energy and 2-RDM error, achieving chemical accuracy across dissociation curves. This framework provides a robust strategy for tomography in many-body quantum simulations.
\end{abstract}

\maketitle

\section{Introduction}

Because measurement of the many-particle wave function scales exponentially with system size, few-particle reduced density matrices (RDMs) are central to the prediction of energies and properties in quantum simulation~\cite{lanyon2010, georgescu2014quantum, mcardle2020, monroe2021, Singh2025}. The energies and properties of many-electron systems can be directly computed from the two-electron reduced density matrix (2-RDM)~\cite{Coleman.2000, MazziottiBook}. In many near-term algorithms, including classical shadow tomography~\cite{aaronson2018shadow, huang2020predicting, zhao2021fermionic, low2022classical}, RDMs are reconstructed from measurement data.  While scalable, such procedures often yield RDMs that violate the constraints required to represent an $N$-electron system---known as the $N$-representability conditions~\cite{coleman1963, erdahl1978representability, harriman1978, Mazziotti2012, Mazziotti2023}---because of statistical and hardware noise. These violations lead to unphysical energies and properties, motivating post-processing methods that restore $N$-representability while remaining consistent with measured data.

A foundation for such correction lies in density-matrix purification, introduced by McWeeny~\cite{mcweeny1960} as a method to restore idempotency of the one-electron RDM and, thereby, enforce $N$-representability. McWeeny purification became the basis for a family of linear-scaling Hartree–Fock and density-functional algorithms~\cite{Goedecker.1999}. Extending this idea to the two-electron RDM (2-RDM), correlated purification~\cite{Mazziotti2002Purification, Pescoller.2025} was developed to restore ensemble $N$-representability conditions beyond the single-particle level by enforcing positivity and ensemble constraints on the 2-RDM.

Here, we extend correlated purification to quantum simulation through a bi-objective variational framework formulated as a semidefinite program. The method minimizes both the many-electron energy and the nuclear norm of the change in the measured 2-RDM. The nuclear norm, often employed in matrix completion~\cite{candes2012exact, Cai.2010}, promotes low-rank, physically meaningful corrections to the 2-RDM, while the energy term acts as a regularization functional that improves energetic accuracy and ground-state purity. This formulation generalizes earlier implementations~\cite{smart2022, rubin2018application} for quantum tomography, which employed a Frobenius norm or omitted energetic information, to provide a more faithful representation of many-body correlations. The present version can be understood as a modification of the variational 2-RDM method~\cite{Mazziotti.20015ua, Nakata.2001, Mazziotti.2002d5q, Mazziotti.2004, Zhao.2004, Cances.2006, Gidofalvi.2008, Shenvi.2010, Verstichel.2011, Baumgratz.2012, Mazziotti.2016sve, Alcoba.2018, Mazziotti.2020z0p, Li.2021, Knight.2022, Mazziotti2023, Gao.2025dgs, Schouten.2025a9e, Piris.2021} in 2-RDM theory through the introduction of a second nuclear-norm objective. Although particularly effective for ground-state reconstruction, the framework can also be applied to excited or nonstationary states by adjusting the relative weights of the energy and error terms.

In applications to fermionic shadow tomography of hydrogen chains on quantum simulators and devices~\cite{low2022classical,wang2025shadow}, this bi-objective correlated purification substantially reduces both energy error and 2-RDM deviation, achieving chemical accuracy across dissociation curves. This approach establishes correlated purification as a scalable, variational, and general strategy for restoring $N$-representability in many-body quantum simulations.


\section{Theory}

Estimating the energy of quantum many-body systems on quantum computers has become a central task in quantum simulations, including but not limited to variational quantum eigensolvers~\cite{Bharti.2022, Head-Marsden.2021, Peruzzo.2014, Cao.2013}. This is typically achieved by expressing the energy as a function of the $p$-particle reduced density matrix ($p$-RDM), followed by its measurement. Consider the Hamiltonian of an $N$-electron system
\begin{equation}
    \hat{H} = \sum_{ijkl} {}^2K^{ij}_{kl}\hat{a}^{\dagger}_i\hat{a}_j^{\dagger}\hat{a}^{}_l\hat{a}^{}_k,
    \label{eq:hamiltonian}
\end{equation}
where ${}^2K$ is the reduced Hamiltonian constructed from one- and two-electron integrals and ${a}^{\dagger}_i$ and ${a}^{}_i$ are the creation and annihilation operators on orbital $i$, respectively. Because the electron interactions are pairwise, the 2-RDM is sufficient to determine the energy and any two-body property~\cite{Coleman.2000, MazziottiBook, coleman1963},
\begin{equation}
    E = \text{Tr}[^2K\; ^2D],
\end{equation}
where
\begin{equation}
    {}^2D^{ij}_{kl} = \bra{\Psi}\hat{a}^{\dagger}_i\hat{a}^{\dagger}_j\hat{a}^{}_l\hat{a}^{}_k\ket{\Psi} ,
\end{equation}
are the elements of the 2-RDM.

Practically, a key challenge in quantum simulations is the realization of accurate and efficient quantum state tomography. Two main issues exist: first, the limited measurement budget, and second, the impact of hardware noise on measurement outcomes. These two factors typically prevent the experimentally measured 2-RDM from corresponding to a physical $N$-electron wave function. Additional post-processing or error mitigation methods are needed to improve the quality of the reconstructed RDM.

For the 2-RDM to be physically meaningful, it must satisfy a set of constraints, which can be expressed as a hierarchy of positivity conditions labeled as the $p$-positivity conditions~\cite{Mazziotti.20015ua, Mazziotti.2006keb, Mazziotti2012, Mazziotti2023}. The most commonly used are the 2-positivity conditions~\cite{Garrod.1964, coleman1963, Mazziotti.2002d5q}, enforcing that the two-particle ($D$), two-hole ($Q$), and particle-hole ($G$) matrices,
\begin{equation}\label{eq:2positivity}
\begin{aligned}
{}^{2}D^{ij}_{kl} &= \bra{\Psi}\hat{a}^{\dagger}_i \hat{a}^{\dagger}_j \hat{a}^{}_l \hat{a}^{}_k\ket{\Psi}, \\
{}^{2}Q^{ij}_{kl} &= \bra{\Psi}\hat{a}^{}_i \hat{a}^{}_j \hat{a}^{\dagger}_l \hat{a}^{\dagger}_k\ket{\Psi}, \\
{}^{2}G^{ij}_{kl} &= \bra{\Psi}\hat{a}^{\dagger}_i \hat{a}^{}_j \hat{a}^{\dagger}_l \hat{a}^{}_k\ket{\Psi},
\end{aligned}
\end{equation}
are positive semidefinite.  A matrix $M$ is positive semidefinite $M \succeq 0$ if and only if its eigenvalues are nonnegative~\cite{VB1996}.  Higher order ($q>2$) constraints including the $T1$ and $T2$ conditions~\cite{erdahl1978representability, Zhao.2004, Mazziotti.2006keb} have also been derived~\cite{Mazziotti2012, Mazziotti2023}. These constraints provide a powerful mechanism for purifying unphysical 2-RDMs with semidefinite programming, both in classical and quantum computing~\cite{foley2012measurement, smart2022, Avdic.2024}.

In this work, we implement a variational correlated purification in which the measured 2-RDM is corrected to satisfy an approximate set of $N$-representability constraints such as the $p$-positivity conditions. One approach is to find the $p$-positive 2-RDM ${}^{2} D_{p}$ that is closest to the experimentally measured 2-RDM $^{2} D_{e}$~\cite{Mazziotti2002Purification, smart2022, rubin2018application}. The notion of closest can be defined through the introduction of a specific distance metric or norm. Because the $p$-positivity conditions require keeping various metric matrices positive semidefinite, the problem with a suitable relaxation of the norm can be formulated as a semidefinite program.  For the 2-positivity conditions, specifically, we have the following program:
\begin{equation}\label{eq:min1}
\begin{aligned}
& \min_{{}^2D_p} &&  \|{}^2D_p - {}^2D_e\|  \\
& \text{such that} && {}^2D_p \succeq 0 \\
& && {}^2Q_p \succeq 0 \\
& && {}^2G_p \succeq 0\\
& & & \text{Tr}({}^2D_p) = N(N-1) \\
& & & {}^2Q_p = f_Q({}^2D_p) \\
& & & {}^2G_p = f_G({}^2D_p) \\
\end{aligned}
\end{equation}
where $f_{Q}$ and $f_{G}$ are linear maps that relate the particle-particle 2-RDM to its hole-hole or particle-hole matrices. In previous work, the distance metric was defined as a relaxation of the Frobenius norm~\cite{smart2022}.

We can make the correlated purification more accurate and efficient by adding a second objective---in the form of the electronic energy as a functional of the 2-RDM---to the optimization. The introduction of the energy term has several key advantages for increasing the fidelity of the estimated 2-RDM, especially at low-shot-budget limit, where data is incomplete or strongly affected by a large variance. First, it prioritizes 2-positive 2-RDMs that minimize not only the normed distance to the experimental 2-RDM but also the energy of the system, which leads to 2-positive 2-RDMs whose energies are much closer to those of the correct, noiseless solution. Second, the term acts as a regularization, removing flat regions of the optimization that can arise from the norm alone.  Finally, it can improve the purity of the 2-RDM's preimage---the $N$-particle density matrix---where the energy minimization biases the optimization towards the boundary of the $N$-representable set of 2-RDMs, particularly for the ground state. Formally, we can write the bi-objective as
\begin{equation}
    \|{}^2D_p - {}^2D_e\| + \frac{1}{w} {\rm Tr}(^{2} K \, ^{2} D),
\end{equation}
in which the weight parameter $w$ controls the strength of the regularization.  For ground states, a small weight $w$ ($w \approx 0.1$) can be highly effective in biasing the result towards the ground state, while for arbitrary quantum states, such as excited states and non-stationary states~\cite{Liebert2021, Benavides-Riveros.2022, wang2023excited}, a much larger $w$ ($w > 1$) is more appropriate.

In addition to the energy regularization, rather than using the Frobenius norm, we introduce a relaxation of the nuclear norm for the distance between the measured and 2-positive 2-RDMs.  Often employed in matrix completion~\cite{candes2012exact, Cai.2010}, the nuclear norm leads to a low-rank, potentially more physical correction to the 2-RDM. We linearize the problem through the introduction of positive semidefinite slack error matrices $E^+$ and $E^-$
\begin{equation}\label{eq:error_blocks}
\begin{aligned}
{}^2D_p - {}^2D_e &= E^+ - E^-, \\
E^+ &\succeq 0, \\
E^- &\succeq 0,
\end{aligned}
\end{equation}
which capture the positive and negative parts of the error matrix. Minimizing the sum of their traces, $\text{Tr}(E^+) + \text{Tr}(E^-)$ is equivalent to minimizing the nuclear norm of the distance matrix, $({}^2D_p - {}^2D_e)$, which favors a low-rank correction~\cite{candes2012exact, recht2010guaranteed}.


The complete semidefinite program for the correlated purification can be expressed as follows
\begin{equation}
\begin{aligned}
& \min_{{}^2D_p, E^+, E^-} & &  \text{Tr}({}^2K {}^2D_p)+ w(\text{Tr}(E^+) + \text{Tr}(E^-)) \label{eq:min_problem} \\
& \text{such that} & & {}^2D_p \succeq 0 \\
& & & {}^2Q_p \succeq 0 \\
& & & {}^2G_p \succeq 0 \\
& & & {}E^+ \succeq 0 \\
& & & {}E^- \succeq 0 \\
& & & \text{Tr}({}^2D_p) = N(N-1) \\
& & & {}^2Q_p = f_Q({}^2D_p) \\
& & & {}^2G_p = f_G({}^2D_p) \\
& & & {}^2D_p - {}^2D_e = E^+ - E^-
\end{aligned}
\end{equation}
The first term in the objective is from the standard variational two-particle reduced density matrix (v2RDM) method~\cite{Mazziotti.20015ua,Nakata.2001,Mazziotti.2002d5q,Mazziotti.2004,Zhao.2004,Cances.2006,Gidofalvi.2008,Shenvi.2010,Verstichel.2011,Baumgratz.2012,Mazziotti.2016sve,Alcoba.2018,Mazziotti.2020z0p,Li.2021,Knight.2022,Mazziotti2023,Gao.2025dgs,Schouten.2025a9e}. The second term enforces data fidelity by minimizing the nuclear norm. The parameter $w$ controls the trade-off between v2RDM ($w \to 0$) and fidelity to the experimental data ($w \to \infty$).


We use the ${}^2D_e$ from fermionic classical shadows (FCS) with particle number symmetry as the experimental starting point and apply the above methodology to obtain purified results. Classical shadows are known to provide polynomial scaling for estimating the 2-RDM. However, their accuracy is limited by both statistical errors and hardware noise, particularly in the low-shot-budget regime. We have implemented FCS as described in Ref.~\cite{low2022classical}. The correlated purification protocol is applicable to any quantum tomography technique that generates a 2-RDM. Moreover, the scheme can be readily extended to restore $N$-representability of any $p$-RDM using the $p$-positivity conditions.

\section{Results}

We benchmark our method on linear hydrogen chains. Fermionic classical shadow data are collected using code interfaced with PySCF~\cite{Sun.2018}, \texttt{ffsim}~\cite{ffsim}, and Qiskit~\cite{Qiskit}. We perform the constrained semidefinite programming with a boundary-point algorithm~\cite{Mazziotti.2011} implemented within the Quantum Chemistry Toolbox in Maple~\cite{maple,rdmchem}. For hydrogen atoms, we use the Slater-type-orbital (STO-3G) basis set~\cite{hehre1969}. Each H$_n$ system's active space consists of $n$ electrons in $n$ orbitals with $\langle {\hat S}_z \rangle = 0$. The first set of benchmarks is obtained from noiseless quantum simulators, where statistical noise is introduced through finite measurement sampling to emulate realistic shot noise. The latter part of this section presents experimental data collected on IBM quantum hardware for comparison\footnote{Both numerical data and a sample computational script are openly available at \url{https://github.com/damazz/CorrelatedPurification}}.

\begin{figure}[t!]
    \centering
    \includegraphics[width=\linewidth]{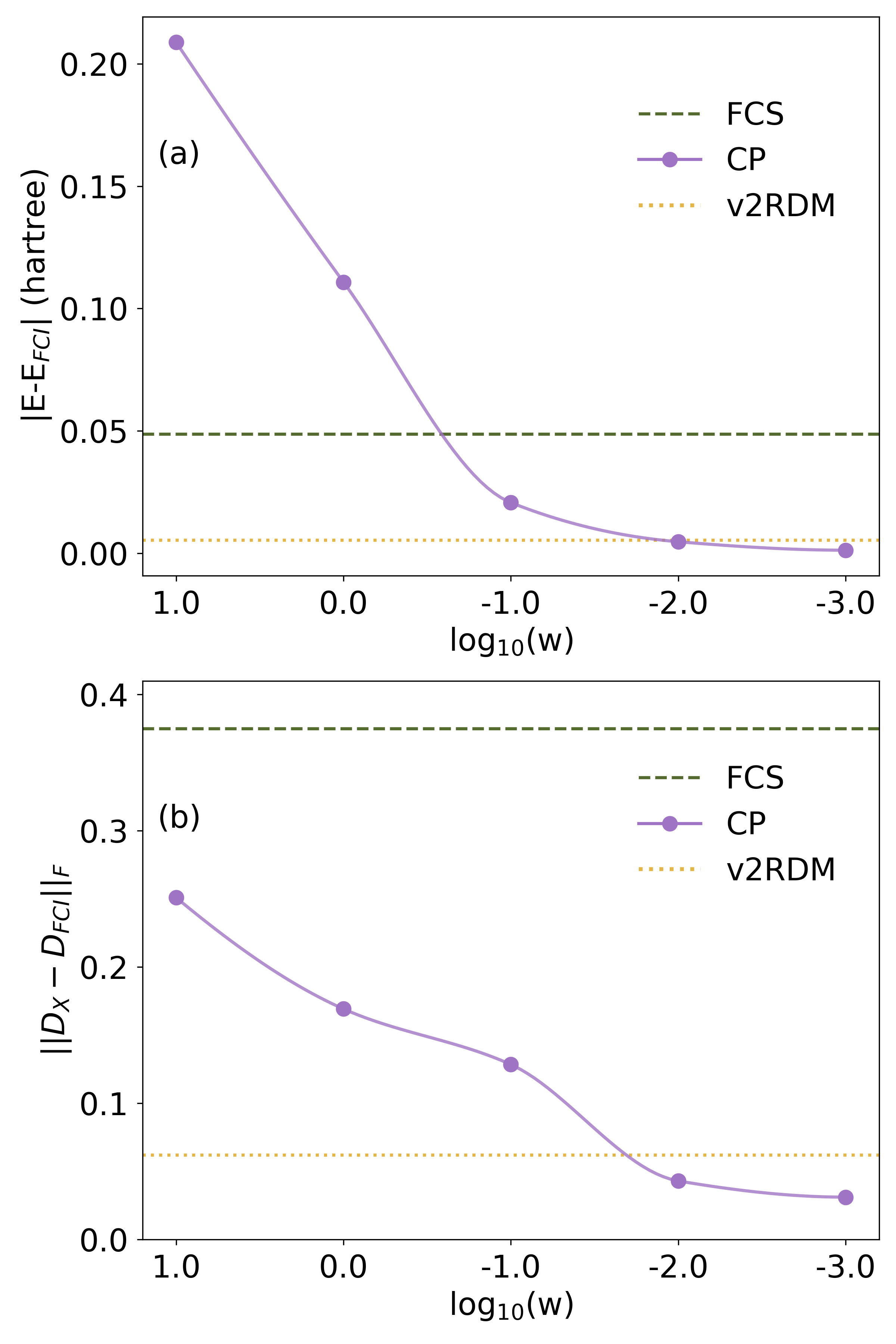}
    \caption{Both (a) absolute energy error and (b) 2-RDM deviation as shown as a function of the weight $w$ for fermionic classical shadows (FCS), the variational two-electron reduced density matrix (v2RDM) method, and correlated purification (CP). For FCS, $10^{5}$ shots are used in the estimation.}

    \label{fig:changew}
\end{figure}

For H$_{4}$, Fig.~\ref{fig:changew} shows the variation of the energy error and 2-RDM deviation as functions of the weight parameter $w$. For small $w$, the optimization in correlated purification (CP) emphasizes the energy term, effectively driving the purified 2-RDM toward the variational minimum. This regime yields markedly smaller energy errors compared with the uncorrected fermionic classical shadows (FCS) under a limited shot budget, demonstrating the strong regularizing effect of the energy minimization. In contrast, for large $w$, the optimization favors the fidelity to the measured 2-RDM, which can limit the improvement if the experimental data are strongly affected by noise. These trends highlight the importance of energy regularization in stabilizing the reconstruction when the measured 2-RDM provides an unreliable reference. Across all $w$, the CP 2-RDM consistently exhibits smaller deviations from the exact result, indicating that the $N$-representability constraints provide a robust correction independent of regularization strength.

\begin{figure}[t!]
    \centering
    \includegraphics[width=\linewidth]{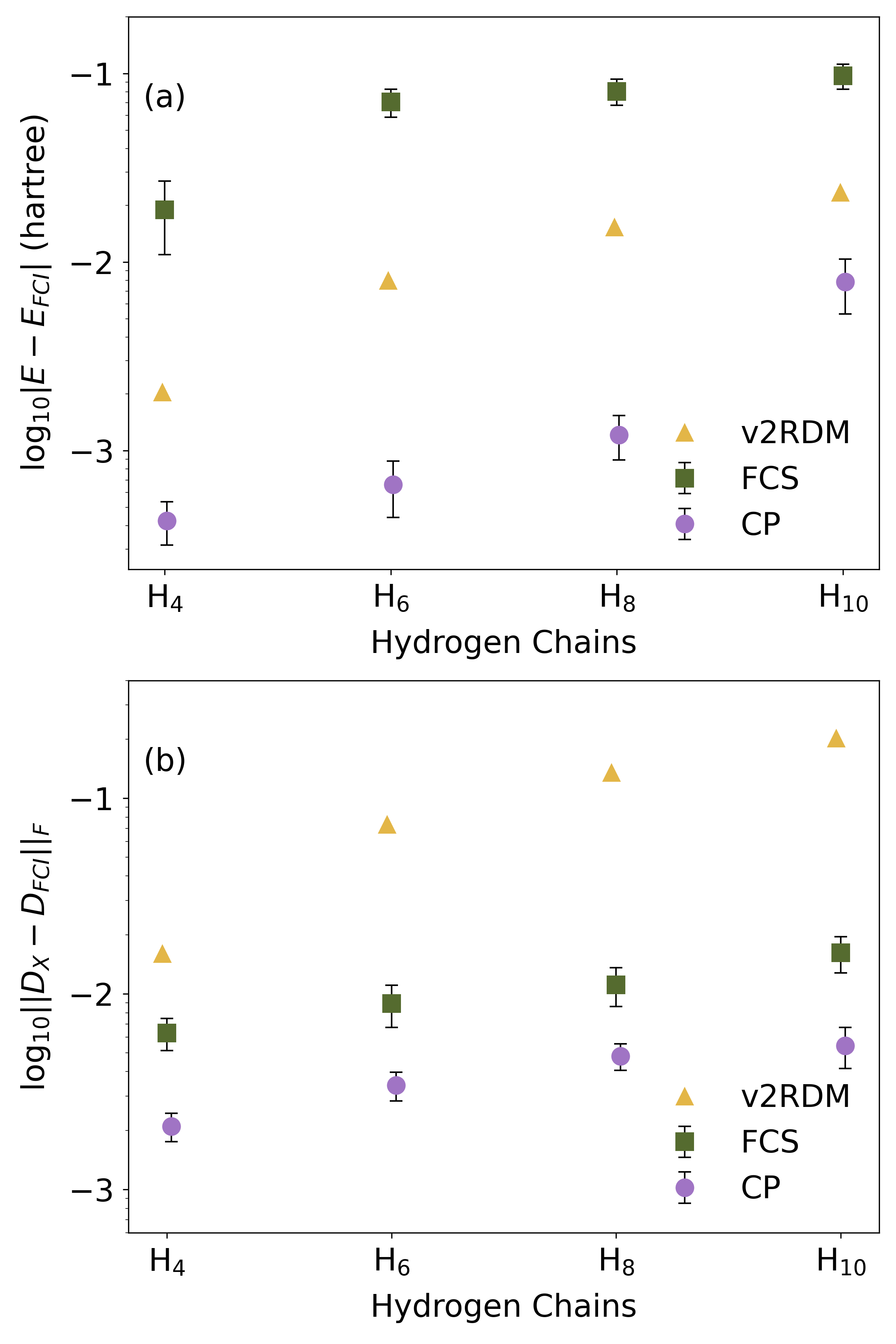}
    \caption{Comparison of error metrics for hydrogen chains H$_{n}$ for $n=4$ through $n=10$ using $10^{6}$ shots and $w = 0.001$. The top panel (a) shows the absolute energy error, and the bottom panel (b) shows the 2-RDM deviation for FCS, v2RDM, and CP. Error bars denote 95\% confidence intervals (approximately 2$\sigma$) computed from these measurements.}

\label{fig:combined_results}
\end{figure}

The optimal choice of $w$ depends both on the quality of the measured data and on the character of the quantum state. For ground states (or approximate ground states) where experimental 2-RDMs are noisy, smaller values of $w$ are preferred, as they leverage the variational energy minimization to enforce physical consistency. For states with weaker ground-state overlap, a larger weight---combined with a higher measurement budget---can provide more stable corrections by preserving information from the measured 2-RDM. Even in the limit $w \to \infty$, CP continues to improve the physical accuracy of the reconstructed 2-RDM relative to the experimental input, proving the intrinsic value of enforcing $N$-representability.

We are particular interested in the scalability of the FCS and our error mitigation protocol. We report the effectiveness of our error mitigation protocol in Figure~\ref{fig:combined_results}, which compares both the absolute energy error and the 2-RDM fidelity for various hydrogen chain molecules. The FCS results consistently exhibit the largest error for energy estimations. As the system size grows, the deviation of v2RDM also emerges on both metrics. The error after purification is lower than that of both the FCS and v2RDM methods across all molecular systems. This robust performance highlights the capability of the purification scheme to accurately reconstruct the 2-RDM, outperforming both baselines instead of interpolating between them.

\begin{figure}[t!]
    \centering
    \includegraphics[width=\linewidth]{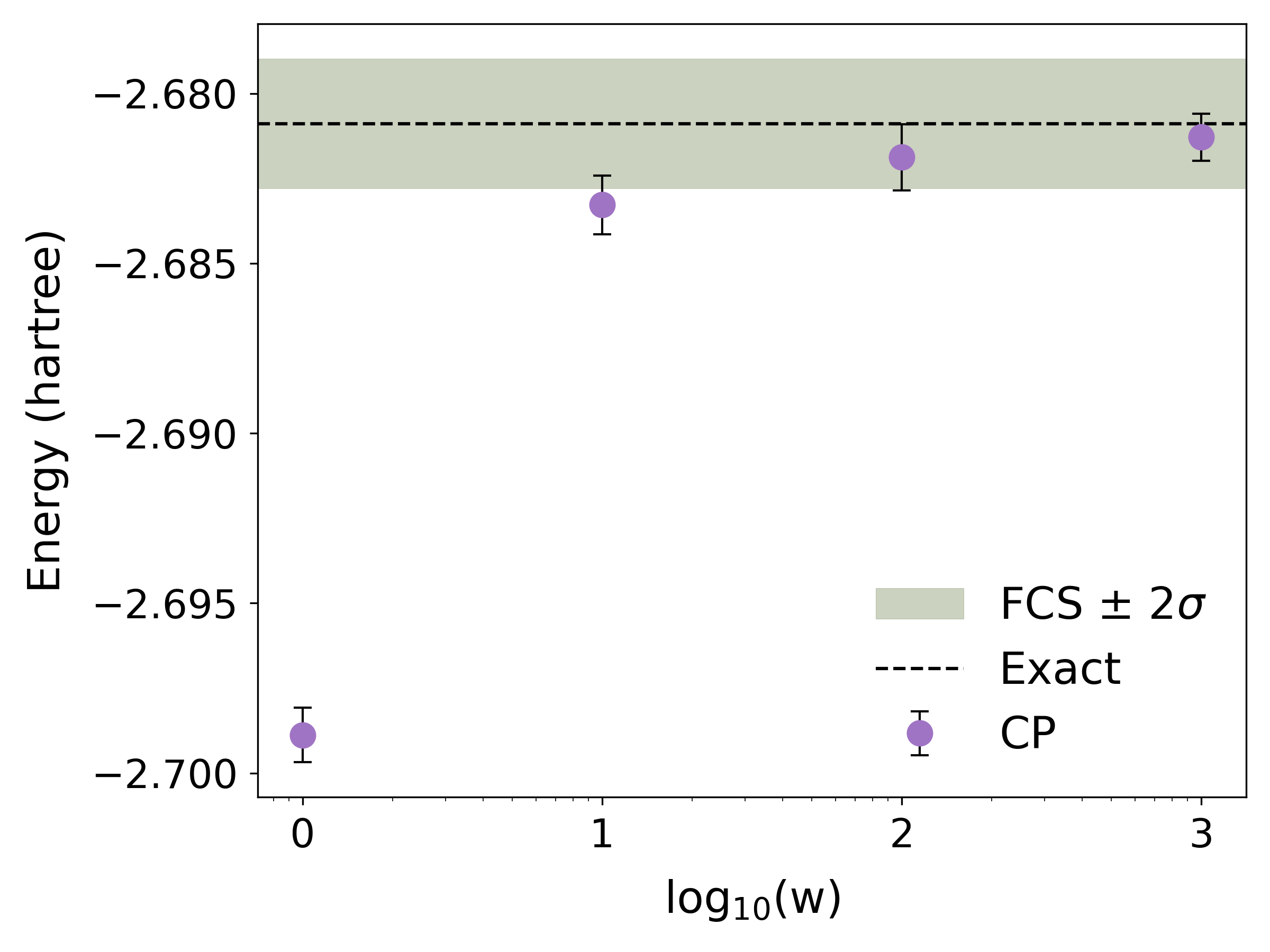}
    \caption{Energy of the 7th excited state of H$_6$ as a function of the weight parameter $w$. Each data point represents an average over 50 independent runs. At larger $w$, CP exhibits a smaller standard deviation than the FCS results, demonstrating improved statistical stability of the energy.}

    \label{fig:excited}
\end{figure}

We demonstrate that excited states can also be captured within our CP framework. One concern with incorporating energy minimization is that constrained optimization might unconditionally favor the ground state. Indeed, this tendency is observed for small $w$, as shown in Fig.~\ref{fig:excited}. Increasing $w$ suppresses the influence of the energy-regularization term, allowing the excited-state features to emerge while maintaining physical consistency. In this regime, CP continues to improve upon the fermionic classical shadow (FCS) results, yielding reduced variance across measurements. This behavior is consistent with our earlier analysis that, in the limit $w \to \infty$, the method still provides a physics-informed correction driven by the $N$-representability constraints. However, we note that the ability to recover missing information under a low-shot budget diminishes at larger $w$, since additional measurements are required to obtain a sufficiently accurate experimental reference, as illustrated in Fig.~\ref{fig:changew}(b).

\begin{figure}[t!]
\centering
\includegraphics[width=\linewidth]{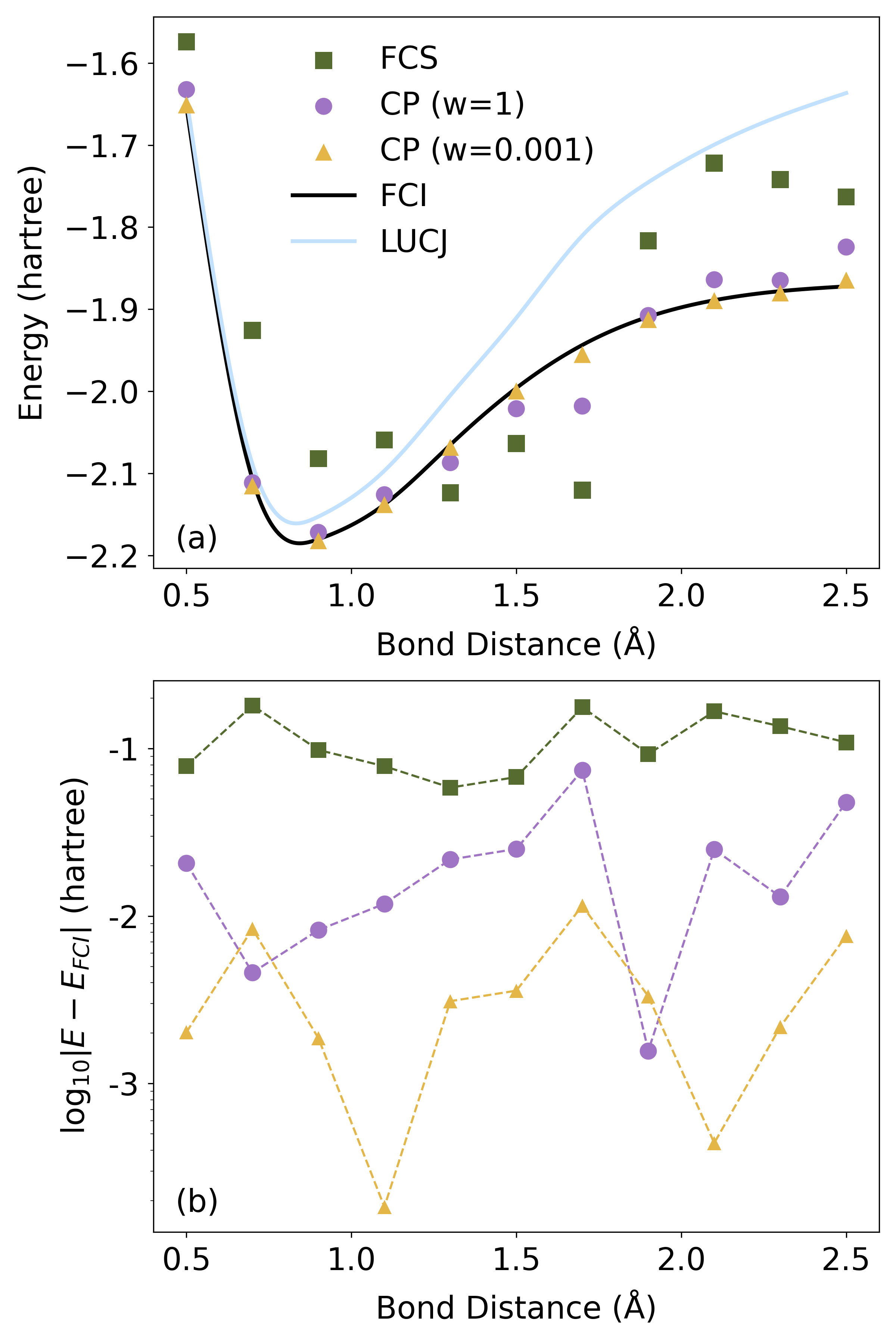}
\caption{Dissociation curve and energy error for the H$_4$ molecule obtained on the IBM \textit{fez} quantum device. (a) Potential energy curves from full configuration interaction (FCI), FCS, and CP ($w = 1$ and $w = 0.001$) as functions of bond distance. (b) Absolute energy error of FCS and CP ($w = 1$ and $w = 0.001$) relative to FCI on a logarithmic scale. CP maintains chemical accuracy across most bond distances, while the FCS results exhibit significantly larger deviations.}

\label{fig:h4_dissociation}
\end{figure}

We next assess the performance of our methods on real quantum hardware. Figure~\ref{fig:h4_dissociation} presents the results for the dissociation of the H$_4$ molecule, obtained from the IBM \textit{Fez} quantum computer. Given the limited shot budget, the FCS results are visibly noisy, with energy errors reaching up to 0.2~hartree. In contrast, CP consistently corrects the noisy data and provides a substantial improvement over FCS across the entire bond-dissociation curve.  Importantly, in the hardware case, because LUCJ differs from FCI, CP is overcoming limitations in both the ansatz and the tomography.  The energy error after purification is comparable with chemical accuracy (1.6~mhartree) for most bond lengths, whereas the uncorrected FCS significantly exceeds this critical limit.

\begin{figure}[t!]
    \centering
    \includegraphics[width=\linewidth]{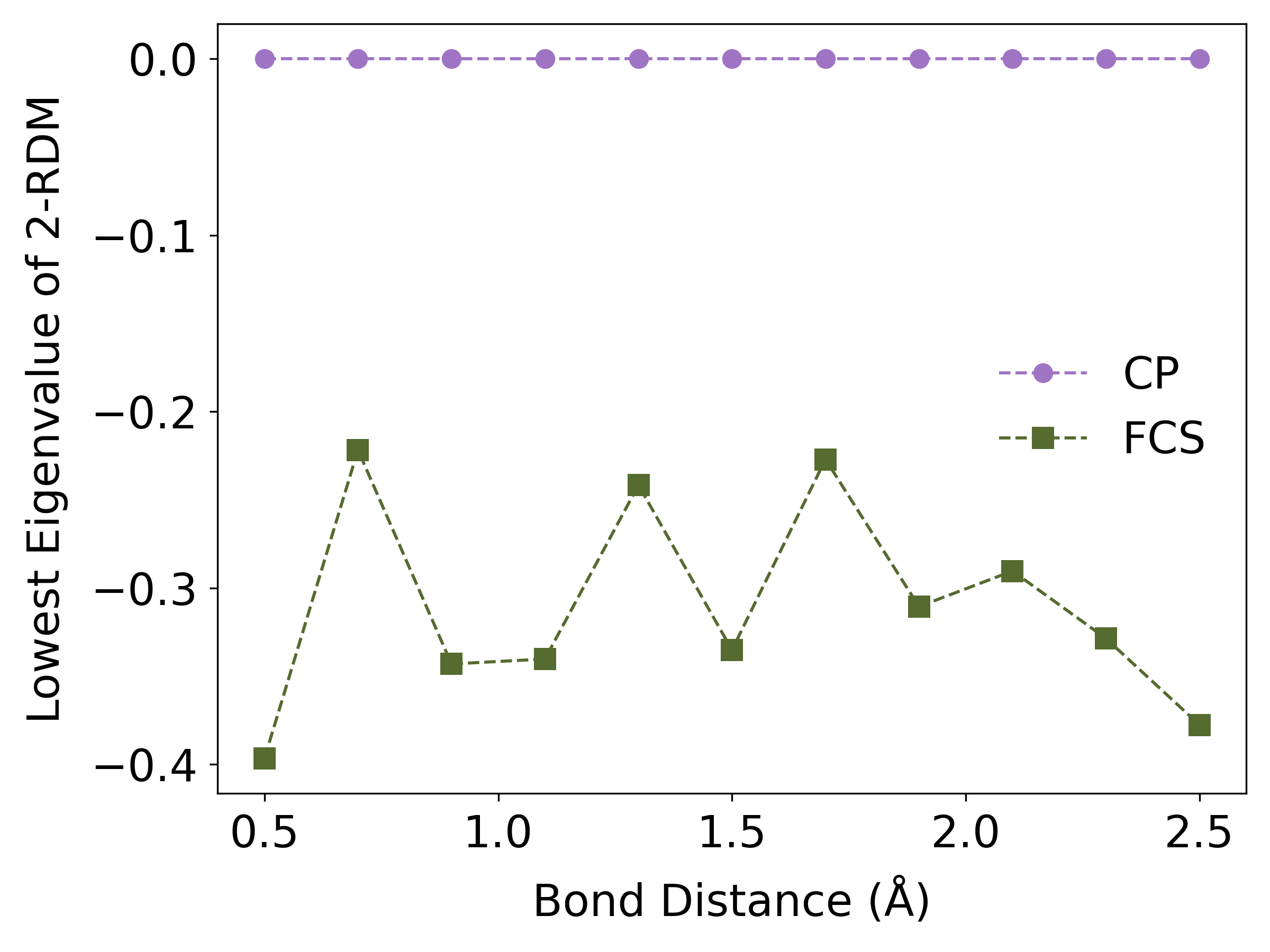}
    \caption{Lowest eigenvalues of the two-electron reduced density matrix (2-RDM) along the H$_4$ dissociation curve shown in Fig.~\ref{fig:h4_dissociation}, obtained from FCS and CP. The FCS results exhibit unphysical negative eigenvalues, while CP enforces positivity under all conditions.  Note that the lowest eigenvalue from CP with any $w$ is effectively zero, and hence, $w=1$ and $w=0.001$ are not shown separately.}

    \label{fig:eigenvalue}
\end{figure}

For each FCS data point, measurements were collected using 10{,}000 random unitaries with one shot per unitary. The mapping employed the Jordan--Wigner transformation~\cite{Jordan.1928}, and the state preparation circuit used the local unitary cluster Jastrow (LUCJ) ansatz with one repetition~\cite{motta2023}. The parameters of the LUCJ ansatz is initialized with t1 and t2 amplitude from coupled clusters singles and doubles and then optimized with respect to energy using \texttt{ffsim} package. Due to hardware noise, including qubit readout errors, some measured bitstrings did not conserve particle occupation number. Such unphysical samples were filtered during post-processing. We note that these artifacts occur only in the real-device experiments and are absent in the noiseless simulators.

The unphysical nature of measured 2-RDM manifests itself through negative eigenvalues in the particle–particle block. For H$_{4}$ Fig.~\ref{fig:eigenvalue} shows that realistic FCS data exhibit this issue due to both statistical and hardware noise. Our constrained optimization strictly enforces the $D$, $Q$, and $G$ conditions, eliminating negative eigenvalues and guaranteeing ensemble $N$-representability for all values of $w$. This enforcement of positivity is essential for recovering physically meaningful RDMs on current quantum hardware.

\section{Discussion and Conclusions}

We have introduced a protocol for restoring the $N$-representability of noisy two-electron reduced density matrices (2-RDMs) obtained from quantum measurements. The method employs a bi-objective optimization that balances two competing goals: minimizing the nuclear-norm deviation from the measured 2-RDM and minimizing the many-electron energy. This dual-objective design ensures that the final 2-RDM remains consistent with experimental data while satisfying the necessary $N$-representability constraints. A tunable parameter $w$ controls the balance between these two objectives, providing flexibility for diverse applications—from correcting low-shot, noisy measurements to refining excited-state calculations.

Benchmarking on 2-RDMs reconstructed from fermionic classical shadows of linear hydrogen chains demonstrates that correlated purification substantially improves accuracy over the uncorrected data. The method remains numerically stable as the system size increases, effectively suppressing error growth in both total energy and 2-RDM. These results show that the approach not only enhances energetic accuracy, but also reconstructs a physically meaningful 2-RDM that better reflects the underlying electronic correlations.

Because the optimization is variational and general, the framework is broadly applicable as a post-processing tool for both classical and quantum algorithms that yield noisy RDMs.  For example, it can be applied to correcting the 2-RDM in time-independent and time-dependent solutions of the contracted Schr{\"o}dinger equation~\cite{Mazziotti.2002, Lackner.2015}. The present formulation can be extended naturally to include higher-order $N$-representability constraints (such as T$_{1}$ and T$_{2}$ conditions) or adaptive weighting strategies in which $w$ is dynamically adjusted according to data quality. By combining correlated purification with the variational 2-RDM framework, this approach provides a scalable, physically grounded error-mitigation strategy that enables reliable quantum simulation of molecular systems and other strongly correlated materials.

\begin{acknowledgments}

We thank Nate Earnest-Noble and Jamie Garcia for valuable discussions. D.A.M gratefully acknowledges support from IBM under the IBM-UChicago Quantum Collaboration. I.A. gratefully acknowledges the NSF Graduate Research Fellowship Program under Grant No. 2140001.

\end{acknowledgments}







\bibliography{references}

\end{document}